\documentclass[prl,twocolumn,showpacs]{revtex4-1}
\usepackage{graphicx}
\usepackage{ulem}
\usepackage{dcolumn}
\usepackage{bm}
\usepackage{natbib}

\begin{document}

\title{Observed Cosmological Redshifts Support Contracting Accelerating Universe}

\author{Branislav Vlahovic}
\email{vlahovic@nccu.edu}
\affiliation{Department  of Physics, North Carolina Central University, 1801 Fayetteville 	Street, Durham, NC 27707 USA.}


\begin{abstract}
\noindent
The main argument that Universe is currently expanding is observed redshift increase by distance. However, this conclusion may not be correct, because cosmological redshift depends only on the scaling factors, the change in the size of the universe during the time of light propagation and is not related to the speed of observer or speed of the object emitting the light.  An observer in expanding universe will measure the same redshift as observer in contracting universe with the same scaling.  This was not taken into account in analysing the SN Ia data related to the universe acceleration. Possibility that universe may contract, but that the observed light is cosmologically redshifted allows for completely different set of cosmological parameters $\Omega_M, \Omega_{\Lambda}$, including the solution $\Omega_M=1, \Omega_{\Lambda}=0$. The contracting and in the same time accelerating universe explains observed deceleration and acceleration in SN Ia data, but also gives significantly larger value for the age of the universe, $t_0 = 24$ Gyr. This allows to reconsider classical cosmological models with $\Lambda =0$. The contracting stage also may explain the observed association of high redshifted quasars to low redshifted galaxies.
\end{abstract}
\maketitle

\section{Cosmological Redshift}

In co-moving Robertson-Walker coordinate system, relation between the redshift $z$, in the frequencies of spectral lines from distant galaxies as compared with their values observed in terrestrial laboratories, and scale factor $a(t)$ is given by
\begin{equation} \label{redshift}
  1+z = a(t_0)/a(t_e),
\end{equation}
where the light emitted with an object at time $t_e$ is observed at the present time $t_0$. Such redshift $z$ is frequently interpreted in terms of the Doppler effect. The reason for that is that for a decreasing or increasing $a(t)$, the proper distance to any co-moving light source decreases or increases with time, so that such sources are approaching us or receding from us. For this reason the galaxies with wavelength shift $z$ are often said to have a cosmological radial velocity $cz$. In series of papers \cite{1R}, in the 1920's, Wirtz and Lundmark showed that Slipher's red shifts, summarized in \cite{Slipher}, increased with the distance and therefore could most easily be understood in terms of a general recession of distant galaxies.

	However there is a problem with interpretation of the cosmological redshift as a Doppler shift. First, the change in wavelength from emission to detection of light does not depend on the rate of change of $a(t)$ at the times of emission or detection, but on the increase of $a(t)$ in the whole period from emission to absorption. We cannot say anything about the velocity of a galaxy at time $t_e$ by measuring redshift $z$, nor we should express the radial galaxy velocity through $cz$. The observed redshift (wavelength stretching) is caused by expansion of space during the entire period of time from emission to observation, not by the velocity of the galaxy at the time $t_e$. The galaxy could have any velocity at the time of emission $t_e$ and we cannot measure it, because we observe just a cumulative effect of wavelength change during the time period $t_0-t_e$.

Cosmological redshift is not related to the speed of the object that emits the light nor the speed of observer. It is also not related to the relative velocity between objects. Our universe can expand or contract, but we will measure the same cosmological redshift. The redshift that will be observed will be given only by the equation (\ref{redshift}) and the scale factors of the observer and emitter.

Because the scaling factor of us as observers is always bigger (it is  by definition $a(t_0)=1)$, than the scaling factor of an object emitting the light $(a(t_e)<1)$, we will always measure the cosmological redshift, and newer cosmological blueshift. This is true even when universe is in a contracting phase, for the present stage of the universe. All object emitting the light, which were at a smaller scale factor during the universe expansion era will have also a smaller scale factor during the universe contracting era, and therefore will be always cosmologically redshited. Our galaxy can approach toward or recede from a object, but in both cases we will measure the same cosmological redshift of that object. The only difference will be due to the proper velocity of our galaxy, which is small on cosmological scale and can be ignored. Cumulative effect of wavelength changes due to the space expansion, is always much bigger than the change of wavelength due to the proper velocities of our galaxy, so that cosmological redshift  $a(t_0)/a(t_e)$ will always dominate over the Doppler redshift, except for some very close objects. This will have as we will see significant implications on the interpretation of the SN Ia data.

The observed redshift does not mean that the universe is currently expanding. We may measure the same amount of the redshift in a contracting universe as in expanding one.  What we observe is only the actual change in the scale factor from the moment of emission of light by an object at $t_e$ to the moment of detection by observer at $t_0$. The redshift does not say anything about the current relative motion between object that emitted the light and observer. For instance, at the present time we may either recede from the surface of last scatter that emitted CMB, we may move toward it, or we may not move at all, but we will be not able to determine the real situation, because in all three cases we will measure the same redshift.

 What we measure is just the change in the scale $a(t)$ from $t_e$ to $t_0$ and not a relative velocity between the objects at $t_e$ and $t_0$. So we may be leaving in an expanding, contracting, or stationary universe, but we cannot determine in which one by just measuring redshift.

\begin{figure}[h]
		\includegraphics[width=7.5cm]{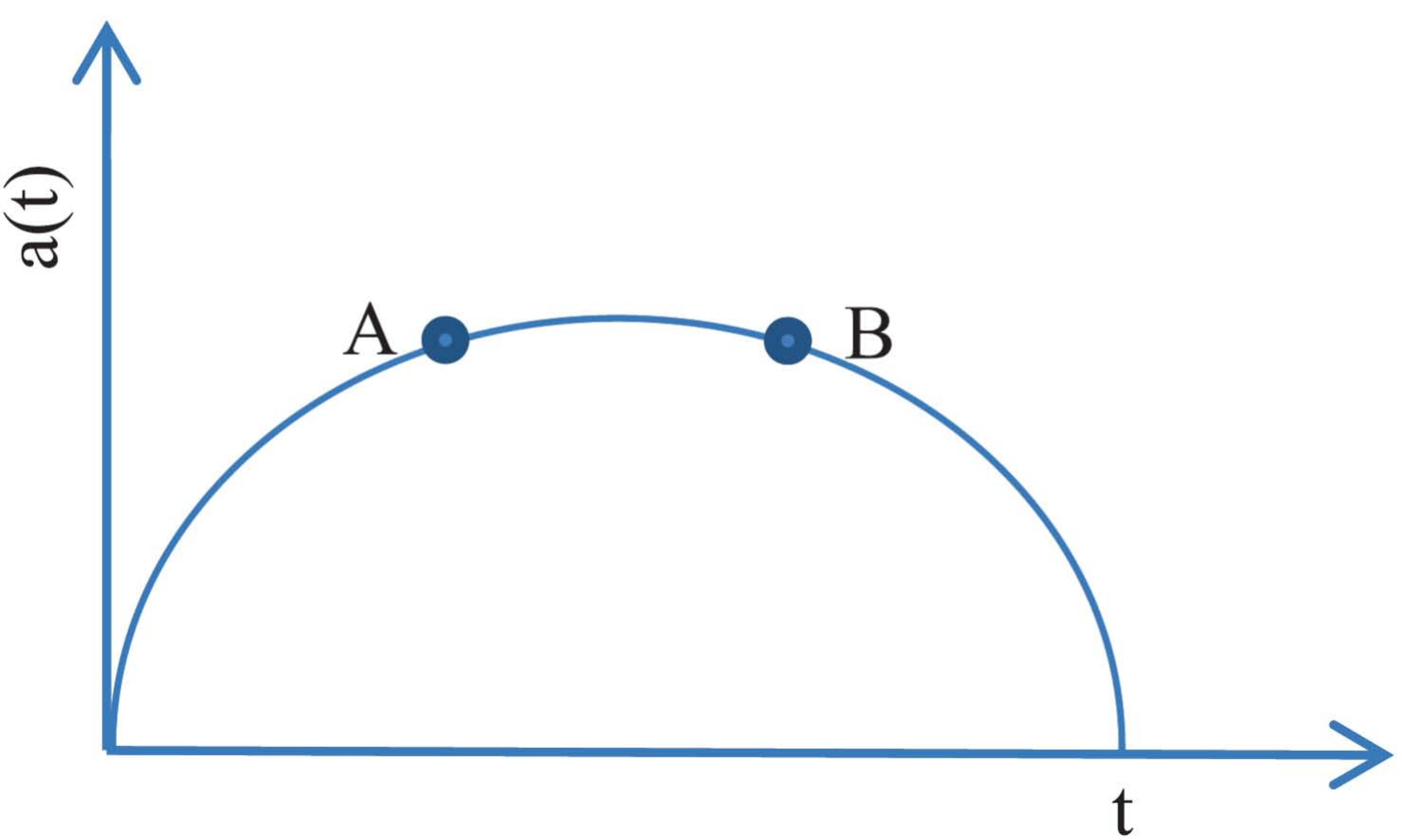}
		\caption{\label{space-expansion}
Scale factor for FRW k=+1 model universe as function of time, with points A and B representing expanding and contracting universe.
}
\end{figure}

To underline this let as consider a closed universe model with k=+1 that initially expands and then contracts. Let consider two points, a point A representing expansion era $H>0$ (where $H\equiv \frac {da/dt} {a(t)})$ and point B representing contracting era $H<0$. Let assume that points A and B are equidistant from the maximum of the expansion (point at $H=0$), i.e. both have the same scaling factor $a$. For that reason both points will have the same scaling ratio $a(t_0)/a(t_e)$ and therefore by equation (\ref{redshift}) will have the same redshift $z$. By measuring redshift we will not be able to conclude if we are in the point A or B, see Fig. \ref{space-expansion}.

\section{Contracting Accelerated universe}

The above make questionable the arguments for reinstating cosmological constant $\Lambda$. The time evolution of the cosmic scale factor and observed redshift depend on the composition of mass-energy in the universe. The universe ordinary matter decelerates the expansion, while exotic form of energy may also significantly affect its dynamics. Observed accelerated expansion \cite{PRS}, \cite{PRS1} in 1998 and assumption that the universe is currently expanding require introduction of dark energy and $\Lambda$. Otherwise current physics cannot explain accelerated universe, which is equivalent to a mass accelerating during climbing up in gravitational potential. We will argue that universe is in the contracting phase (regardless of redshift, which is not relevant) and for that reason it actually must accelerate, because it corresponds to an object falling in gravitational field.

We will show that observed acceleration obtained by magnitude-redshift relation, comparing the apparent magnitudes of high and low redshift SN Ia supernovae, can be explained by assuming that universe is in contracting stage. This may be an indication that the universe is at the present in the point B rather than in the point A.

The high-redshift supernovae, as compared with low-redshift supernovae, appear almost 0.15 mag ($\approx$ 15\% in flux) fainter than expected in a low mass density $(\Omega_M=0.2)$ universe without a cosmological constant, which was the base for established universe acceleration.

The universe acceleration can be expressed through deceleration parameter, which is defined by the scaling factor $a(t)$ as:
\begin{equation} \label{qasa}
  q \equiv - \frac {\mbox{\"a}a} {\mbox{\.a}^2}.
\end{equation}
We can also express deceleration, $q$, through Hubble expansion parameter:
\begin{equation} \label{qasH}
  \frac {\mbox{\.H}} {H^2}=-(1-q).
\end{equation}
  All forms of matter yield a deceleration parameter $q \geq -(1+q)$, except the case that violates all the energy conditions. Thus, in any expanding universe the Hubble parameter should decrease and the expansion should decelerate.

 The Friedman-Robertson-Walker magnitude redshift relation can be expressed as \cite{P97}:
 \begin{equation} \label{magnitude}
  m_B^{effective} \equiv M_B+5logD_L(z;\Omega_M,\Omega_{\Lambda}),
\end{equation}
            where $M_B\equiv -5logH_0+25$ is "Hubble constant free" B- band absolute magnitude at maximum of a SN Ia with width $s=1$ and $D_L\equiv H_0d_L$ is the "Hubble constant free" luminosity distance.
     So, the distances estimated from SN Ia are derived from the luminosity distance relation
     \begin{equation} \label{d_l}
  d_L = \left( \frac{L} {4\pi F}\right)^{\frac {1} {2}},
\end{equation}
where $L$ and $F$ are the SN Ia's intrinsic luminosity and observed flux, respectively.
Luminosity distance $d_L$ at a given redshift $z$ can be further expressed through cosmological parameters, the mass density, ${\Omega}_M$, and dark energy or cosmological constant, ${\Omega}_{\Lambda}$:
   \begin{equation} \label{d_l1}
 d_L =  cH_0^{-1}(1+z)|{\Omega}_k|^{-1/2}S\{{\Omega}_k|^{1/2}
   \end{equation}
 $\hspace{0.5in}$ $ \times\int_{0}^{z}dz[(1+z)^2(1 + {\Omega}_Mz)- z(2+z){\Omega}_{\Lambda}]^{-1/2}\}$,

where ${\Omega}_k = 1-{\Omega}_M - {\Omega}_{\Lambda}$, and $S$ is $sinh$ for ${\Omega}_k \geq 0$ and sin for ${\Omega}_k \leq 0$.
After correcting observed peak luminosity of SNe Ia data by Multi-Color Light Curve Shape (MLCS) method (which employs the observed correlation between light curve shape, luminosity, and color) obtained are precise light curve distance. The cosmological parameters ${\Omega}_M$ and ${\Omega}_{\Lambda}$ which give the best fit to the Hubble diagram and also satisfy other constrains, CMB data, the dynamics of clusters of galaxies, and the evolution of clusters, are selected together with associated likelihood.

High-redshift SNe Ia are observed to be dimmer than expected in an empty universe (i.e.,
 $\Omega_M = 0$) with no cosmological constant, which suggest an eternally expanding universe which is accelerated by dark energy.  Mass density in the universe, $\Omega_M > 0$, exacerbates this problem, requiring
even more dark energy.

 The problem with the analysis performed by all groups analyzing SNe Ia data is that the allowed range for the parameters ${\Omega}_M$ and ${\Omega}_{\Lambda}$ did not included an contracting universe. Considered are only parameters that satisfy $H>0$. The reason for that is the assumption that observed redshift means expansion, which is as it was discussed earlier not correct.

  The basic idea behind the method is that for an object of known brightness, the fainter the object the farther away it is and the further back in time. One can consequently argue that if the universe is in the contracting stage, that it will imply smaller luminosity distance and an increase in luminosity, which will be opposite than observed. However, that is a naive and not correct assumption. Even in contracting universe, the lookback time will increase and the luminosity distance will increase. The propagation of light must follow world line geodesics through 4-dimensional spacetime. If our galaxy is moving from point A to pint B on Fig. \ref{space-expansion}, the light must follow the same world line.

  To illustrate that luminosity distance will increase even during the universe contracting era let us just consider earlier example where universe is moving from point A to point B. Both points have the same scaling factor, but they are separated by time. That will reflect in the luminosity distance between points A and B. The luminosity distance between point A and point B will be not zero $d_L \neq 0$ even both points have the same $z$ value $z_A - z_B = 0$.

If the option with physically allowed parameters: $\ddot{a}(t_0)>0$, $H<0$ is allowed in \cite{PRS} and \cite{PRS1} or more recently in \cite{Giostri} then solution ${\Omega}_M =1$ and ${\Omega}_{\Lambda}=0$
will be possible.

  To demonstrate that it is enough to place an observer in point $B$ and to place point $B$ at about $z=0.8$ relative to the point of the maximum expansion or $H=0$. This will reproduce observed acceleration in SN Ia data with $z < 0.8$, and observed deceleration in SN Ia with $z > 0.8$. This will be in accordance with recent estimation \cite{Giostri}
  that acceleration happened at $z_a =0.64_{-0.07}^{+0.13}$ and that there was a transition period between deceleration and acceleration  that corresponds to a time interval $\gamma = 0.36_{-0,17}^{+0.11}$. This make distance between points $A$ and $B$ to be $z = 1.64$, which corresponds to about 10 Gyr.

\section{Dynamical Age of the Universe }

  In no decelerating universe, such as an empty universe the age of the universe is simple $t_0 = H_0^{-1}$. For an universe that has more complex cosmology the age of the universe can be expressed as:
     \begin{equation} \label{time}
 t_0(H_0,\Omega_M,\Omega_{\Lambda}) = H_0^{-1} \int_{0}^{\inf} (1+z)^{-1}
 \end{equation}
$\hspace{.5in} \times [(1+z)^2(1+\Omega_Mz)-z(2+z)\Omega_{\Lambda}]^{-1/2}dz$,

 where integration is form the present $z=0$ to the beginning $z=\infty$.
 This gives for $t_0 = 14.2 \pm 1.5$ Gyr \cite{PRS1}.

 This value is actually too low, if compared with more recent but still elusive estimates for the ages of the globular clusters, between 12 and 20 Gyr, for example \cite{Berg}, \cite {Gratton}, \cite{Chaboyer}. The age of the globular cluster is thus about the same as the age of the universe. Nothing less problematic are high redshift objects, as for instance: galaxy A1689-zD1 at the redshift $z = 7.6$ \cite{Bradley}, gamma ray burst, GRB 090423, which had a redshift of $z=8.2$ \cite{Salvaterra}, and UDFy-38135539 redshift of $z=8.6$, which is the highest confirmed spectroscopic redshift of a galaxy \cite{Lehnert}, it corresponds to just 600 million years after the Big-Bang. So short time to the Big-Bang could be problematic to explain formation of that galaxy.

 However, if our galaxy is at the point $B$ instead at the point $A$, then our universe is as discussed above older for about 10 Gyr. It removes any problem related to the age of the universe. It also allows to consider a typical cosmological model with $\Lambda = 0$, since having our galaxy at the present in the point $B$ creates enough time to allow to reach the expansion maximum.

 We should not dilute this article with addressing other not necessary related subjects. However, it is hard to resist to mention that position of our galaxy in the point $B$ and contracting stage of the universe may also allow to explain observable indications that quasars of generally larger redshifts are associated with larger galaxies of much lower redshift \cite{Arp 1990}.
 For closed universe in FRW  metric
\begin{equation} \label{metric}
                      ds^2= c^2dt^2-a^2(t)[d\chi^2 + sin^2\chi(d\theta^2 + Sin^2\theta d\phi^2)]
\end{equation}
where $\chi=\int cdt/a(t)$, which is by definition conformal time coordinate or arc parameter time $\eta$.
 Substituting for $ds^2 = 0$ and assuming that our galaxy is located at $\chi=0$ and that observed quasar is at   point $\chi_{Q}$, the light will reach us when
\begin{equation} \label{eta}
                      \eta_0 - \eta_Q = \int_{0}^{t_0} \frac{cdt}{a(t)} = \frac{c}{a_0}\int_{0}^{z}\frac {dz'}{H(z')}.
\end{equation}
 We can only see high redshifted QSO as it was billion years ago, we cannot see how it looks at the present time. If it exist today, it may be for instance in a form of a galaxy, which developed from such quasar. However, that galaxy if exist is separate spacelike and not observable. However, in an contracting universe, if the amount of the shrinking is adequate, such previously not visible galaxy can be visible now, due to the contraction of the space. So, it may happen that in the same time we can see highly redshifted QSO in it original form, how it was billion of years ago, and in its new form, in the form of galaxy to which it transformed. The images of these two objects may overlap and also the new galaxy may create multiple images of the QSO by gravitational lensing.

\section{Conclusion}

The cosmological redshift depends only on the scaling factors of the universe at the time when light was emitted and observed, it is not related to the speed of the object emitting the light or speed of the observer. This allows to observe cosmologically redshifted light in a contracting universe. When this possibility is included in analysing SN Ia data it leads to a new set of cosmological parameters that do not require dark energy, i.e. allows solutions with $\Omega_M=1, \Omega_{\Lambda}=0$. Assuming that current universe started to contract and accelerate at $\sim z = 0.8$ reproduces SN Ia observable data. It also leads to an older universe with age $t_0 = 24$ Gyr.
The contracting stage of the universe also allows to explain association of the high redshifted quasars and low redshifted galaxies.

\begin{acknowledgments}
I would like to thank S. Matinyan, I. Filikhin, and A. Soldi for useful discussions. This work is supported by NSF award HRD-0833184 and NASA grant NNX09AV07A.
\end{acknowledgments}

\end{document}